\documentclass[aps,prl,preprint,showpacs,showkeys,groupedaddress,amsmath,amssymb]{revtex4}

\usepackage{graphicx}
\usepackage{dcolumn}
\usepackage{bm}

\begin{document}


\title{Spin wave resonances in  La$_{0.7}$Sr$_{0.3}$MnO$_{3}$  films: measurement of spin wave stiffness and anisotropy field}

\author{ M. Golosovsky\footnote{Permanent address: the Racah Institute of Physics, the Hebrew University of Jerusalem, 91904, Jerusalem, Israel},  P. Monod, }%
\affiliation{Laboratoire de Physique du Solide, ESPCI, 10 rue Vauquelin, 75231 Paris Cedex 05, France}
\author{P.K. Muduli, R.C. Budhani}%
\affiliation{Department of Physics,Indian Institute of Technology, Kanpur, 208016, India\\}
\date{\today} 

\begin{abstract} 
We studied magnetic field dependent microwave absorption  in epitaxial La$_{0.7}$Sr$_{0.3}$MnO$_{3}$ films  using an X-band Bruker ESR spectrometer. By analyzing angular and temperature dependence of the ferromagnetic  and spin-wave resonances we determine spin-wave stiffness and anisotropy field. The spin-wave stiffness as found from the spectrum of the standing spin-wave resonances  in thin films is in fair agreement with the results of inelastic neutron scattering studies on a single crystal of the same composition [Vasiliu-Doloc et al., J. Appl. Phys. \textbf{83}, 7343 (1998)].
\end{abstract}
\maketitle

\section{introduction}
Spin-wave stiffness, $D$, is an important parameter of  magnetic materials that  characterizes the magnon dispersion law, $\omega=Dk^{2}$. In the mean-field approximation
it is directly related to exchange integral $J$, namely, 
\begin{equation}
D=\frac{2JSa^{2}}{\hbar}
\label{exchange}
\end{equation}
where $\gamma$ is the gyromagnetic ratio, $a$ is the lattice constant, and $S$ is the spin.\cite{Kittel}  Inelastic neutron scattering studies of the spin-wave stiffness in manganite single crystals  \cite{Martin,Lynn,Vasiliu,Vasiliu2,Vasiliu3,Zhang} revealed important information on the nature of their ferromagnetic  transition. Since this issue is  under debate \cite{Salamon,Zhang}  it  is important to study spin-wave stiffness  more closely, in particular, to measure $D$ in thin films. It is also important to explore complementary  techniques.  In principle, spin-wave excitations can be measured   using  magneto-optical Kerr effect or Brillouin light scattering \cite{Patton} although existing studies of manganites by these methods \cite{Azevedo,Talbayev} do not focus on spin waves.

Spin-wave stiffness has been traditionally measured by the microwave absorption technique: using a fixed frequency ESR spectrometer \cite{Patton,Farle} or broadband techniques.\cite{Grishin}
Several groups reported fixed frequency microwave studies of standing spin-wave resonances (SWR) in  La$_{0.67}$Ba$_{0.33}$MnO$_{3}$ (Ref. \cite{Lofland}),  and La$_{0.7}$Mn$_{1.3}$O$_{3}$ (Refs. \cite{Dyakonov2,Dyakonov3,Aleshk2}) thin films. Although spin-wave excitation at  microwave frequencies was observed  in La$_{0.75}$Sr$_{0.11}$Ca$_{0.14}$MnO$_{3}$ (Ref. \cite{Yin}), La$_{0.7}$Ca$_{0.3}$MnO$_{3}$ (Ref. \cite{Shames}) and La$_{0.7}$Sr$_{0.3}$MnO$_{3}$ (Ref. \cite{Lyfar}) thin films,  it was not studied in detail.  

In contrast to inelastic neutron scattering  that (i) measures travelling spin waves with large wavevectors, $k\sim$ 1-0.1\AA$^{-1}$ and (ii) requires large single crystals;  the microwave absorption technique measures spectrum of standing spin-wave resonances with small wavevectors, $k\sim 10^{-2}-10^{-3}$\AA$^{-1}$ and operates mostly with thin films.  The recent overview  of spin-wave excitations in manganites \cite{Aleshk1} compares thin-film microwave absorption measurements to inelastic neutron scattering studies of single crystals. It turns out that there is no manganite compound that was studied simultaneously by both techniques. Our present work fills this gap. We study  spin-wave spectrum  in epitaxial La$_{0.7}$Sr$_{0.3}$MnO$_{3}$ films of different thicknesses and on different substrates by the microwave absorption technique at 9.4 GHz and compare our data to the inelastic neutron scattering studies on single crystals of the same composition.

\section{Standing spin-wave resonances in a thin ferromagnetic film}
Consider a thin ferromagnetic film with an "easy-plane" magnetic anisotropy. Magnetic field is oriented at oblique angle  $\Psi$ with respect to the film normal. Orientation of magnetization, $\Theta$, is determined by the interplay between the external field, $H$, and the perpendicular anisotropy field, $H_{a}$, and is found from the following equation:\cite{Farle,Gurevich} 
\begin{equation}
H_{a}\sin \Theta \cos\Theta=H\sin({\Theta-\Psi})
\label{angle}
\end{equation}
where in-plane anisotropy has been neglected. In the presence of the microwave magnetic field with frequency $\omega$, whereas $h_{mw}\perp H$, the spin-wave resonances (SWR) are excited. The  resonance field $H$ is found from the following condition:\cite{Farle,Gurevich}
\begin{eqnarray}
\omega^{2}=
[\gamma H\cos(\Theta-\Psi)-\gamma H_{a}\cos^{2}\Theta+Dk^{2}]\nonumber\\
\times[\gamma H\cos(\Theta-\Psi)-\gamma H_{a}\cos2\Theta+Dk^{2}]
\label{spin-wave}
\end{eqnarray}
If surface spins are completely pinned or completely unpinned,  then
\begin{equation}
k=\pi n/d
\label{resonance}
\end{equation}
where $d$ is the film thickness and $n$ is an integer. The uniform FMR mode corresponds to $n=0$, while the spin-wave resonances  correspond to $n\neq$0.  For the perpendicular orientation ($\Psi=\Theta=0$)  Eq. (\ref{spin-wave}) yields
\begin{equation}
H_{n}-H_{0}=H_{a}-\frac{\pi^{2} n^{2}}{\gamma d^{2}}D
\label{sw}
\end{equation}
where $H_{0}=\frac{\omega}{\gamma}$ and $\omega$ is the microwave frequency. To determine $D$, one measures microwave absorption in dependence of magnetic field and notices a sequence of resonances. By analyzing their spectrum using Eq. (\ref{sw}) one identifies mode numbers $n$. The slope of the linear dependence, $H_{n}$ vs $n^2$, yields $D$, while the intercept with the $H$-axis  yields $H_{a}$.
To enable such procedure the film thickness should lie in certain limits,
\begin{equation}
\delta>>d>>\pi\left[\frac{D}{\gamma(H_{0} +H_{a})}\right]^{1/2} 
\label{thickness}
\end{equation}
where $\delta$ is the skin-depth. Indeed, for efficient excitation of the spin-wave resonances  the film thickness should be smaller than the skin-depth. On the other hand, the film should be  thick enough to support several  resonances.

\section{Experimental}
Our experiments were performed with a bipolar $X$-band  Bruker ESR spectrometer, a  $TE_{102}$ resonant cavity, and an Oxford cryostat.  We studied  La$_{0.7}$Sr$_{0.3}$MnO$_{3}$  films ($d=$ 50, 100, 150 and 200 nm) grown on the (001) SrTiO$_{3}$  substrate (STO), and La$_{0.67}$Sr$_{0.33}$MnO$_{3}$ films ($d=$ 50 and 150 nm) on the NdGaO$_{3}$  substrate (NGO). The samples  were fabricated by the pulsed laser deposition \cite{samples} and were cut to small mm-size pieces in order to keep the reasonable value of the cavity $Q$-factor. We measured magnetization and resistivity of these films by SQUID magnetometry and four-point technique, correspondingly. The skin-depth at 9.4 GHz  estimated from our resistivity measurements \cite{Musa}  - 22 $\mu$m at 295 K and 5 $\mu$m at 50 K- considerably exceeds the film thickness. 

The film uniformity could be estimated from the FMR spectrum.  A uniform film is characterized by a narrow FMR peak, corresponding to a well-defined anisotropy field, while a nonuniform film usually exhibits a broad FMR peak indicating wide spread of the anisotropy fields. Most part of our samples demonstrated narrow FMR peaks at ambient temperature, although some samples showed several narrow FMR peaks corresponding to discrete values of anisotropy field. 

\section{Experimental results}
\subsection{La$_{0.7}$Sr$_{0.3}$MnO$_{3}$ films on the SrTiO$_{3}$ substrate}

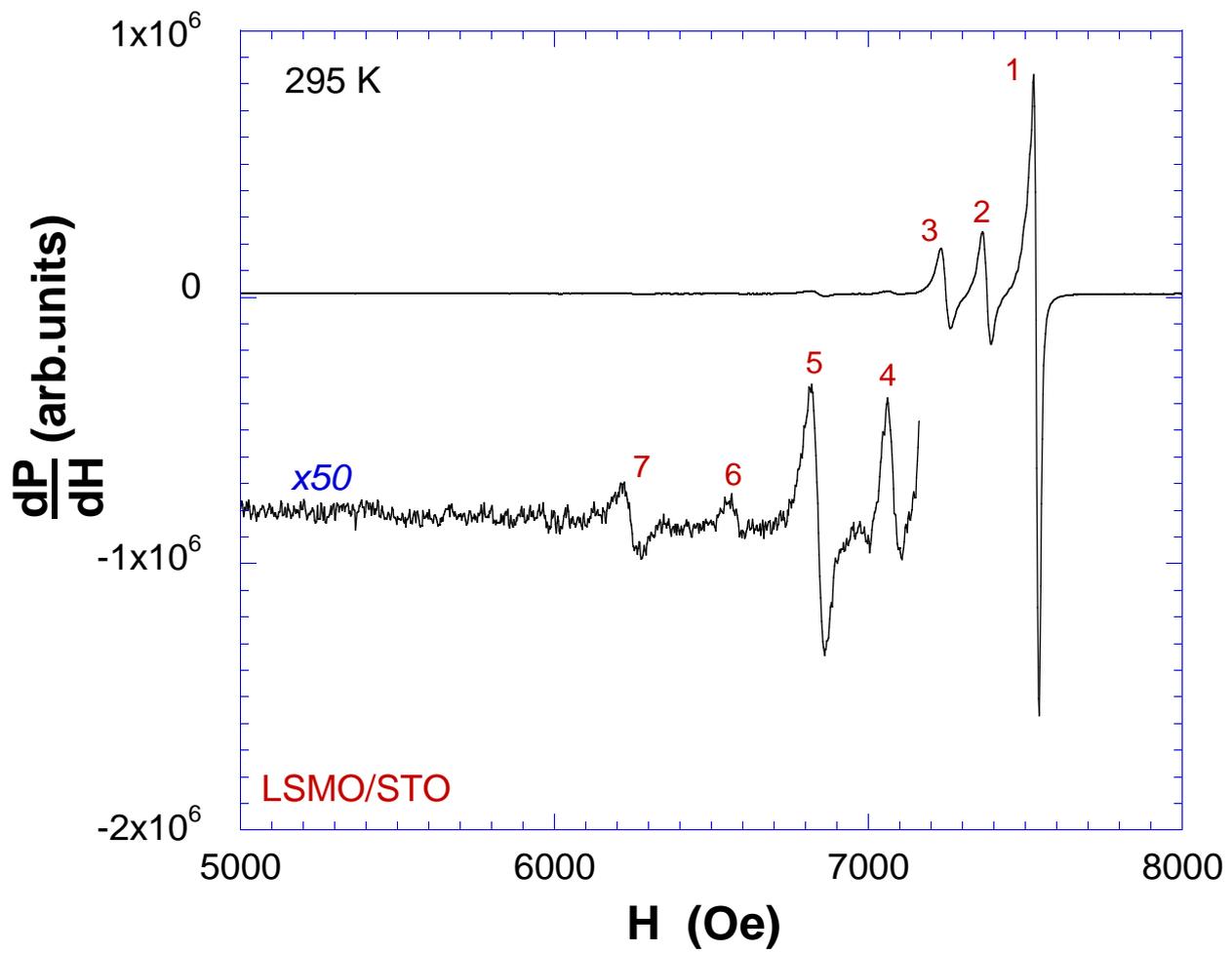
\begin{figure}[ht]
\caption{Absorption derivative spectrum for a 200 nm thick La$_{0.7}$Sr$_{0.3}$MnO$_{3}$ film on the SrTiO$_{3}$ substrate in the perpendicular magnetic field. The mode number is shown at each peak.  The microwave frequency is 9.4 GHz, the incident  power is  $P_{mw}=$ 0.2 mW (-30 dB), the modulation field is 10 Oe.}
\label{fig:LSMO55-H}
\end{figure}

Figure \ref{fig:LSMO55-H} shows microwave absorption derivative in the perpendicular field for a 200 nm thick La$_{0.7}$Sr$_{0.3}$MnO$_{3}$ film on the SrTiO$_{3}$ substrate. We observe a series of slightly asymmetric  narrow peaks. The asymmetry arises from the coupling to the dielectric resonances in the STO substrate and was observed in other studies as well. \cite{Lyfar,Musa} The angular dependence of the resonant fields (not shown here) follows Eq. (\ref{sw}), hence  we attribute these peaks to the spin-wave resonances. We cut the original 5$\times$ 5 mm$^2$ film to five pieces with the size of  1$\times$ 0.5 mm$^2$ and all these pieces  demonstrated fairly identical spectra. The  very fact that we observe sharp and reproducible  resonances  proves that the film is highly uniform. Another indication of the high quality of the film is the narrow peak-to-peak linewidth (16 Oe at ambient temperature) and low coercive field (11-16 Oe at 295 K).

The mode numbers were established as follows. We assigned consecutive  numbers to the peaks in Fig. \ref{fig:LSMO55-H} and checked  whether linear dependence, $H_{n}\propto n^{2}$, predicted by Eq. (\ref{sw}), holds. The best correspondence to Eq. (\ref{spin-wave}) was achieved for the sequence  $n=$1,2,3.. or $n=$0,2,3... There is some ambiguity in whether the strongest peak corresponds to $n=0$, to $n=1$ or to their sum,  since the splitting between the $n=0$ and the $n=1$ modes as predicted by Eq. (\ref{sw}), is  only  25 Oe  and this is comparable to the linewidth.

Figure \ref{fig:linewidth-k} shows SWR intensities and linewidths. The linewidth increases almost linearly with  mode number, while intensity decreases. Odd modes have generally higher intensities than even modes (see also Fig. \ref{fig:LSMO55-H}). 

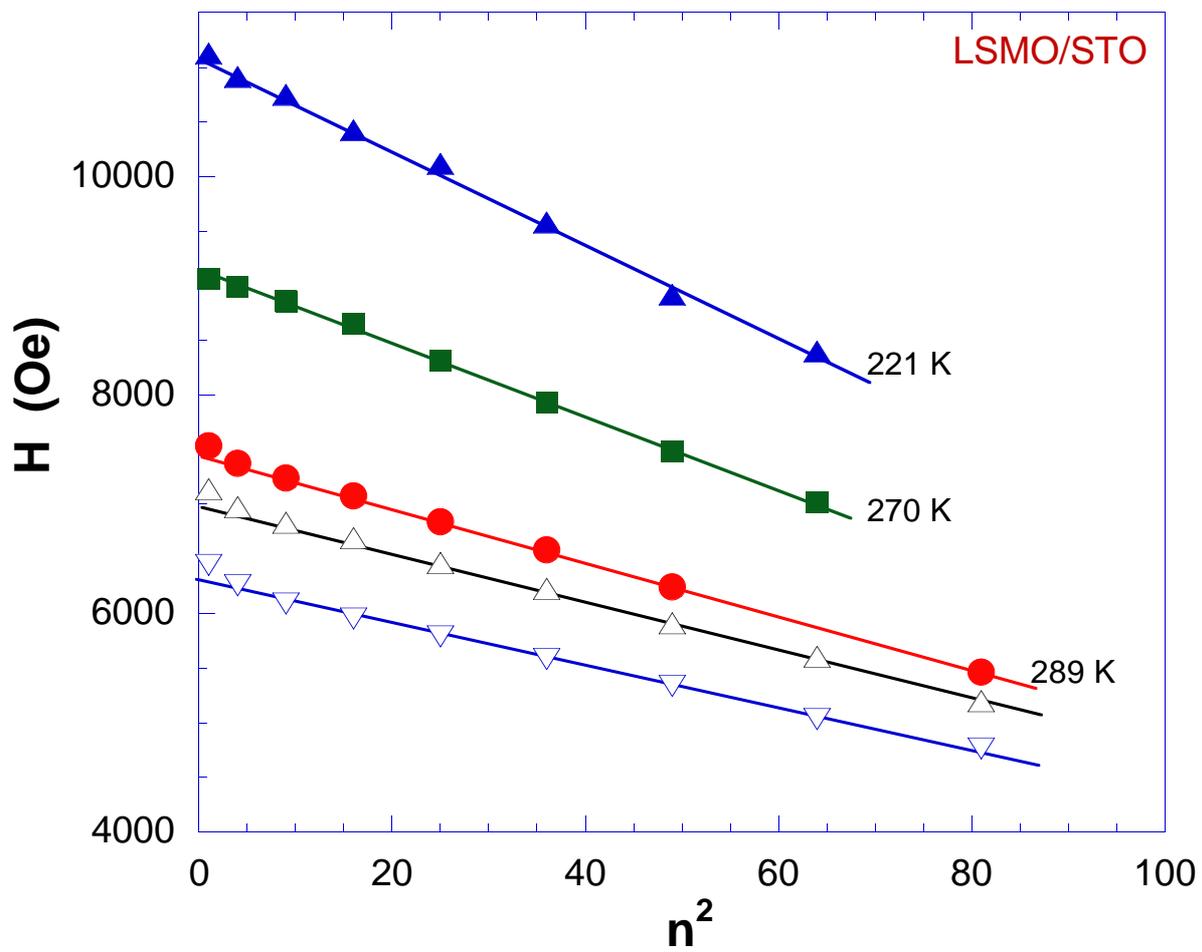
\begin{figure}[ht]
\caption{The resonance field of different modes for the film shown  in Fig.\ref{fig:LSMO55-H}.  Filled symbols indicate the data measured at fixed and known temperatures and at low microwave power, $P_{mw}=0.2$ mW. Open symbols indicate the data measured at ambient temperature and increased microwave power (62.5 mW and 200 mW, correspondingly). Here, the sample temperature is higher than the ambient temperature due to self-heating. The lines show  $H_{n}\propto n^{2}$ approximation.}
\label{fig:LSMO55-harm}
\end{figure}

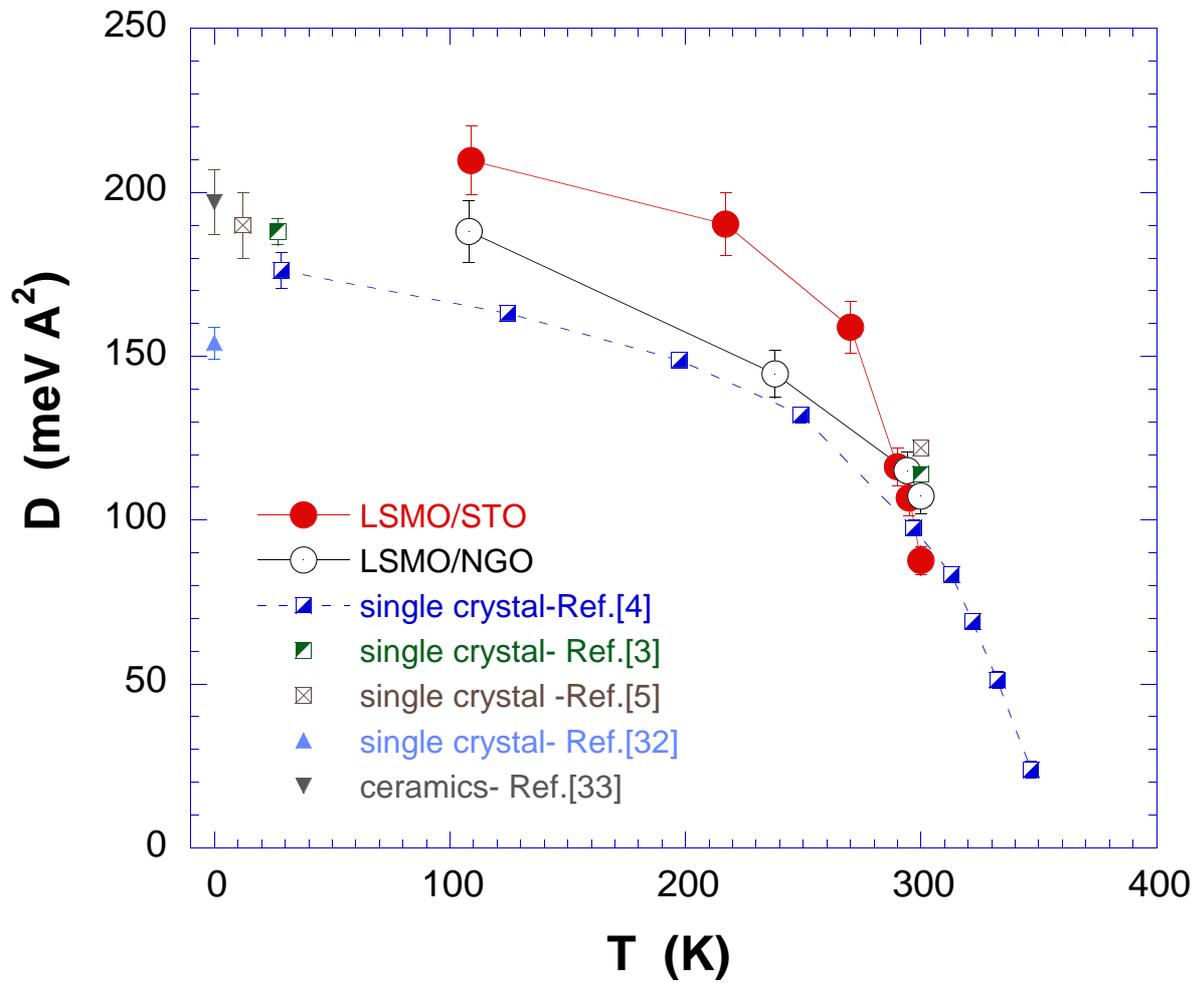
\begin{figure}[ht]
\caption{Temperature dependence of the spin-wave stiffness $D(T)$. The circles stand for our microwave measurements on thin films on substrate. The squares show inelastic neutron scattering data  for the La$_{0.7}$Sr$_{0.3}$MnO$_{3}$ single crystals. The triangles stand for the $D^{Bloch}(T=0)$ estimated from the temperature dependence of magnetization and $T^{\frac{3}{2}}$-Bloch law. The lines are the guide to the eye.}
\label{fig:D-total}
\end{figure}

Figure \ref{fig:LSMO55-harm} shows dependence of  the resonance fields  on the mode number.  The higher-order modes obey quadratic dependence, $H_{n}\propto n^2$, while the modes with low $n$ show tendency to linear spacing that is quite common for the films with surface pinning. \cite{Phillips} We exclude from our analysis  the first two modes that should be most strongly affected by  surface pinning and  determine spin-wave stiffness from the slope of  $H_{n}$ vs $n^2$ dependencies  using  Eq. (\ref{sw}). The results are shown in Fig. \ref{fig:D-total}. 

To find the perpendicular anisotropy field we extrapolate the  $H_{n}$ vs $n^{2}$ dependencies  to $n=0$. Equation (\ref{anisotropy}) and Fig. \ref{fig:LSMO55-harm} yield $H_{a}=$0.4 T  at 295 K and $H_{a}=$1 T  at 4.2 K.  To analyze these values we note that the perpendicular anisotropy field of a thin film, 
\begin{equation}
H_{a}=H_{demag}+H_{cryst}+H_{stress},
\label{anisotropy}
\end{equation}
consists of demagnetizing field- $4\pi M$, crystalline anisotropy- $H_{cryst}$, and stress-induced anisotropy- $H_{stress}$. The crystalline anisotropy of La$_{0.7}$Sr$_{0.3}$MnO$_{3}$ is very small \cite{Suzuki,Ziese} and does not exceed 0.03 T.\cite{Lofland2} Demagnetization field as estimated from magnetization \cite{Suzuki,Ziese} is  $H_{demag}=$0.74 T at 4.2 K. Since the lattice mismatch between La$_{0.7}$Sr$_{0.3}$MnO$_{3}$ and SrTiO$_{3}$ is only 1.4 $\%$  and the film is sufficiently thick, the $H_{stress}$ is not high and achieves considerable magnitude only at low temperatures.\cite{Ranno}  Equation (\ref{anisotropy}) yields the stress anisotropy field, $H_{stress}=$ 0.26 T  at 4.2 K.  This means that even at low temperatures the demagnetization field is the dominant contribution to the anisotropy field. This is consistent with other measurements. Indeed, magnetization studies of the Ref. \cite{Ziese} for the films of comparable thickness found that $H_{demag}=0.8H_{a}$ at 10 K, while  Ref. \cite{Suzuki} found that  $H_{demag}=0.95 H_{a}$ at 295 K.

\subsection{La$_{0.67}$Sr$_{0.33}$MnO$_{3}$ films on the NdGaO$_{3}$ substrate}
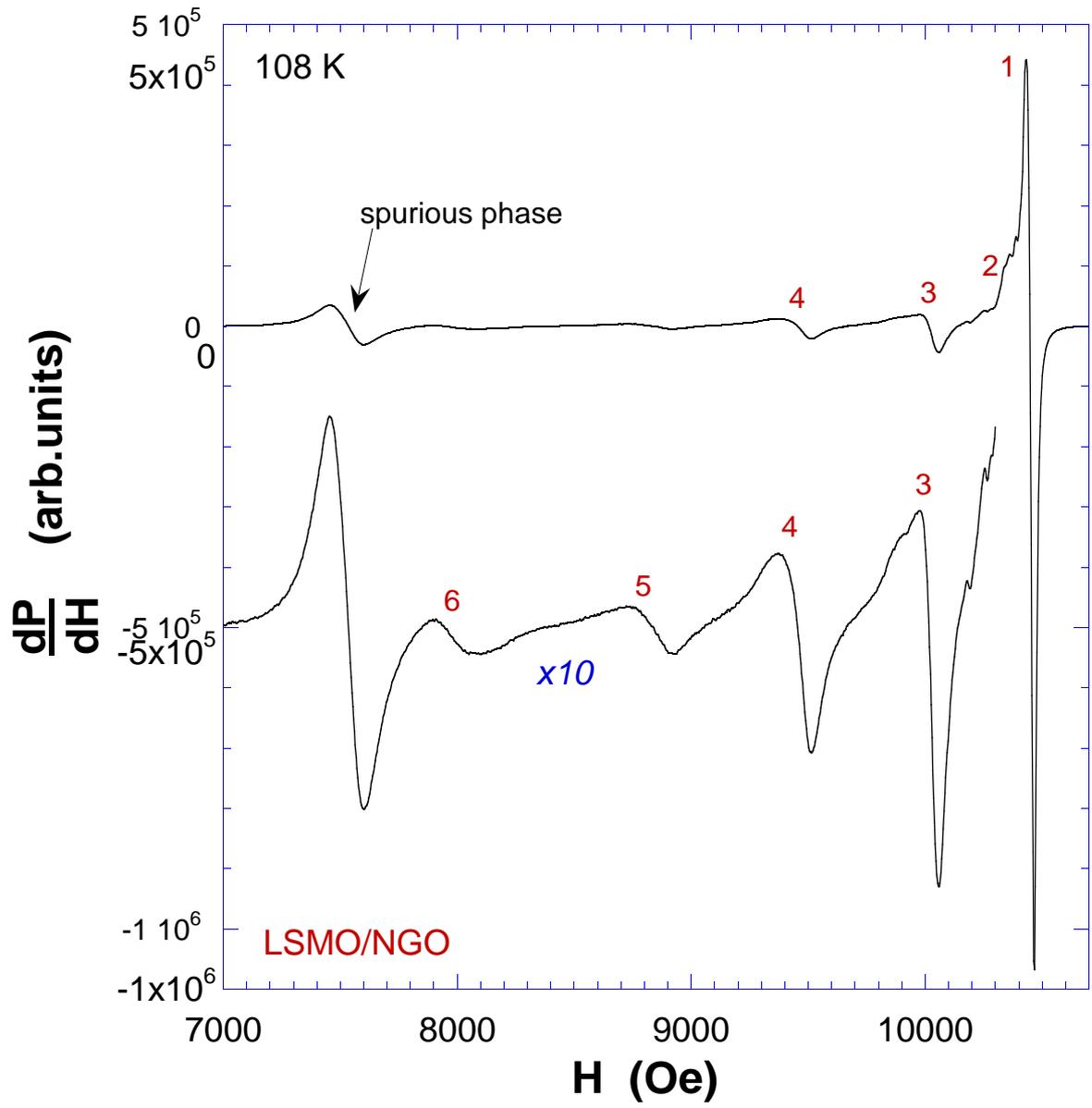
\begin{figure}[ht]
\caption{Absorption derivative spectrum for a 150 nm thick La$_{0.67}$Sr$_{0.33}$MnO$_{3}$ film on the NdGaO$_{3}$ substrate. The mode number is shown at each peak.  A strong peak at $H=$7470 Oe  originates from the ferromagnetic resonance in a region of the film with the different anisotropy field.}
\label{fig:NGO-H}
\end{figure}

Figure \ref{fig:NGO-H} shows microwave absorption spectrum  for a 150 nm thick La$_{0.67}$Sr$_{0.33}$MnO$_{3}$ film on the NdGaO$_{3}$ substrate at 108 K. We observe a strong and narrow peak at 10445 Oe and a series of low-field satellites.  The  peak-to-peak linewidth  of the dominant resonance is very small (12 Oe at ambient temperature and 33 Oe at 108 K) and this proves high quality of the film. The coercive field at ambient temperature is only 4 Oe. We cut the film to several pieces and they showed consistent spectra. 

To identify the resonances we measured  temperature and angular dependencies of the resonant field and came to conclusion that the peaks designated with integer numbers in Fig. \ref{fig:NGO-H} are spin-wave satellites of the peak at 10445 Oe, while the strong peak at $H=$ 7470 Oe does not belong to this series. Indeed, the angular dependencies of resonant field of the numbered peaks are very similar (not shown here) and different from that for the peak at 7470 Oe. The same is true with respect to the temperature dependencies. We attribute the peak at 7470 Oe to the region with a different discrete value of the anisotropy field and exclude it from the subsequent analysis. 

Figure \ref{fig:NGO-harm} shows that the  resonant fields of higher modes follow $H_{n}\propto n^2$ dependence.  The slope of this dependence (the first mode excluded) yields the spin-wave stiffness $D$. The results are plotted in  Fig. \ref{fig:D-total}. 
The same data yield the perpendicular anisotropy field. We find $H_{a}=$0.73 T at 4.2 K. This is almost equal to the demagnetizing field, $4\pi M=$0.74 T, and means that the stress-induced  anisotropy is negligible here, as expected for the  lattice-matched substrate. The linewidth (see inset) steadily increases with $n$. 

\subsection{Other films}
We  observed spin-wave resonances in several films with different thickness. In a very thin ($d=$ 50 nm)  La$_{0.67}$Sr$_{0.33}$MnO$_{3}$ film  on SrTiO$_{3}$ we observed only one spin-wave resonance which was displaced down by 0.23 T from the dominant FMR resonance [while Eq. (\ref{sw}) predicts 0.08 T]. Such pronounced displacement is obviously related  to strong surface pinning, hence it is not possible to measure $D$ there. In several 150 nm thick La$_{0.67}$Sr$_{0.33}$MnO$_{3}$ films  on SrTiO$_{3}$ we observed  two overlapping series of spin-wave resonances, so that determination of $D$ here was ambiguous.  Ref. \cite{Aleshk2} observed such split spin-wave resonances in a 350 nm thick La$_{0.7}$Mn$_{1.3}$O$_{3}$  film on  LaAlO$_{3}$ substrate and showed that the splitting disappears after annealing in oxygen. Following Ref. \cite{Aleshk2} we studied  the effect of oxygen annealing on the SWR spectra in our films. Contrary to  Ref. \cite{Aleshk2}  we found that oxygen annealing at different temperatures  from 600$^{0}$C to 900$^{0}$C does not ameliorate the SWR spectrum in our films but introduces additional splitting. In particular, the resonant peaks in annealed films split into many narrow lines with the spacing of $\sim$15 Oe. The difference between our results and those of Ref. \cite{Aleshk2}  is probably related to the fact that we operate with the film of different composition and on different substrate.

\section{Discussion}
\subsubsection{Zero-temperature spin-wave stiffness} Figure \ref{fig:D-total} compares the spin-wave stiffness found in our microwave absorption studies of La$_{0.7}$Sr$_{0.3}$Mn$_{0.7}$O$_{3}$ thin films to inelastic neutron scattering measurements  on single crystals of the same composition.\cite{Martin, Vasiliu,Vasiliu2} Consider first the limit $T=0$. Our measurements for two films of different thickness yield $D_{0}^{film}=$210 meV \AA$^2$ while  the corresponding neutron scattering data show lower values, $D_{0}^{cryst}=$ 188  (Ref. \cite{Martin}), 170 (Ref. \cite {Lynn}), 176 (Ref. \cite{Vasiliu}), and 190 meV \AA$^2$ (Ref. \cite{Vasiliu2}). The discrepancy between $D_{0}^{film}$ and $D_{0}^{cryst}$ may arise from the fact that  few atomic layers adjacent to the film-substrate interface are nonmagnetic. In LSMO this "dead layer"  may be up to 5 nm thick,\cite{twolayers,NMR} hence "magnetic" thickness that appears in Eq. (\ref{sw}) is smaller than the nominal thickness. The correction for the "dead layer" can bring down $D_{0}^{film}$ by $~5\%$.

The spin-wave stiffness at $T=0$ can be also estimated from the  temperature dependence of magnetization and the $T^{\frac{3}{2}}$-Bloch law (see also Ref. \cite{Rosov}). For  the La$_{0.7}$Sr$_{0.3}$MnO$_{3}$ single crystals this yields $D_{0}^{Bloch}=$154  (Ref. \cite{Smolyaninova}),  while for ceramics of the same composition, $D_{0}^{Bloch}=$197 meV \AA$^2$ (Ref. \cite{Budak}). 

\subsubsection{Temperature-dependence} Consider now the temperature dependence of the spin-wave stiffness. Figure \ref{fig:D-total} shows that the results for two films with different thicknesses and on different substrates (one with tensile stress-STO and another with a weak compressive stress-NGO) are very close, as expected for intrinsic property.   At ambient temperature, we find for these two films   $D^{film}=$104 and 114 meV \AA$^2$, correspondingly. This is almost identical to the single crystal data at ambient  temperature- $D^{cryst}=$114 (Ref. \cite{Martin}) and 100 meV \AA$^2$ (Ref. \cite{Vasiliu}). However, one should take into account the difference in $T_{C}$ as well.  

\subsubsection{Comparison between the samples} 
In order to compare data for the samples with different $T_{C}$ we  consider  $D(T)$ vs $M(T)$ plots where the temperature is an implicit variable. A similar plot was used earlier by Ref. \cite{Rossing} to compare spin-wave stiffness of the Fe-Cr alloys  with different composition and different $T_{C}$. The rationale behind such plot is the mean-field expression -Eq. (\ref {exchange})- that in fact relates the spin-wave stiffness to magnetization, $M=g\mu_{B}S/V_{a}$. Here, $\mu_{B}$ is the Bohr magneton, $g$ is the $g$-factor, and $V_{a}\propto a^3$ is the atomic volume.  Equation (\ref{exchange}) yields direct proportionality, $D\propto JMa$. Strictly speaking, this proportionality is expected only for the Heisenberg model and at $T=$0.  However, since $M$ and $D$ both go to zero at $T_{C}$, while $J$, in general, varies continuously across the ferromagnetic transition; then we expect that the $D\propto M$ proportionality holds up to $T_{C}$. Indeed, as the Ref. \cite{Kobler} shows, empirical data on many ferromagnetic compounds suggest that $D$ and $M$ have the same temperature dependence  in the whole range from $T=$ 0 to $T=T_{C}$.

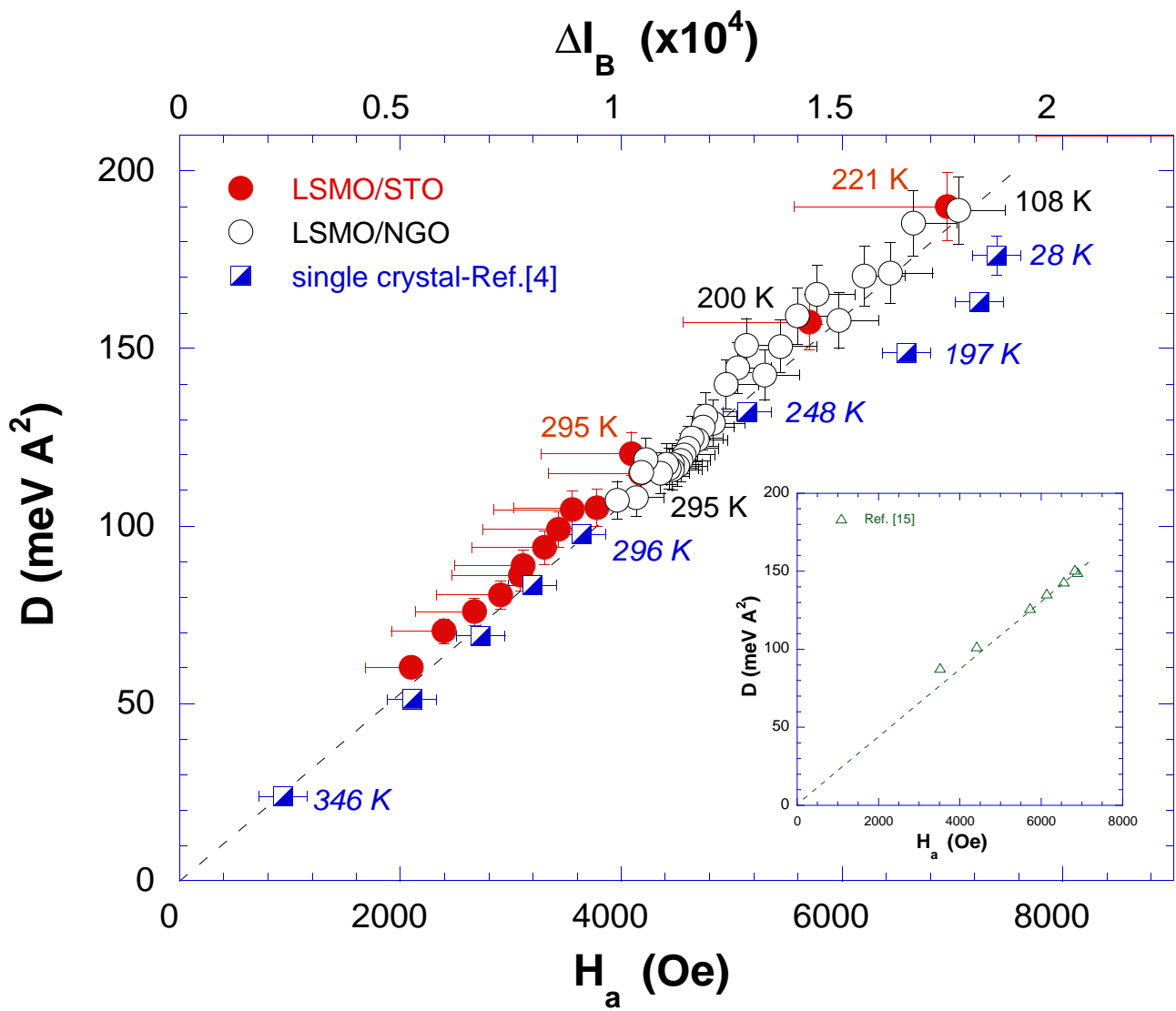
\begin{figure}[ht]
\caption{Spin-wave stiffness  versus magnetization at varying temperature. The circles stand for our microwave measurements on thin films. As a measure of magnetization  we take the perpendicular anisotropy field, $H_{a}$ (lower horizontal scale).  The squares stand for the single crystal  data measured by the  inelastic neutron scattering  technique (Ref. \cite{Vasiliu}). As a measure of magnetization in this case we take  the integrated intensity of the electronic Bragg peak (upper horizontal scale). The dashed line shows linear approximation. The inset shows  similar $D(M)$ dependence for the La$_{0.7}$Mn$_{1.3}$O$_{3}$ film as extracted from the  microwave absorption measurements of Ref. \cite{Dyakonov2,Dyakonov3}.}
\label{fig:M-D-all}
\end{figure}

To effectuate this approach  we plot $D$ vs $M$ (Fig.  \ref{fig:M-D-all}) where $D$ and $M$ are measured at the same temperature.\cite{temperature}  We find $D$  from the microwave absorption spectra, while the magnetization is estimated indirectly, from the anisotropy field, $H_{a}$. Indeed, since the dominant contribution to $H_{a}$ in our films is the demagnetizing field $4\pi M$ (especially for the films on NGO), hence the anisotropy field is a measure of magnetization. 

Figure \ref{fig:M-D-all} plots  the spin-wave stiffness versus anisotropy field where the temperature is an implicit parameter.  It also plots the corresponding  single crystal neutron-scattering data of Ref. \cite{Vasiliu} where magnetization was estimated from the intensity of the electronic Bragg peak.  The upper horizontal scale in Fig. \ref{fig:M-D-all} was  chosen in such a way that the low-temperature limit of the electronic Bragg peak intensity ($\Delta I_{B}=1.85\times 10^{4}$) corresponds to saturation magnetization of La$_{0.7}$Sr$_{0.3}$MnO$_{3}$, i.e. $4\pi M=$ 0.74 T. The horizontal error bars in our thin film measurements take into account the possible difference between the anisotropy field and magnetization arising from the stress anisotropy field [Eq. (\ref{anisotropy})].  We  assume extreme values of $H_{stress}/4\pi M$=-0.06 and 0.2 for the films on NGO and STO, correspondingly.

 We observe that above $T=$295 K the data for both our films and for the single crystal collapse. The resulting $D(M)$ dependence is quaisilinear that indicates the same critical indices of $D$ and $M$ at $T_{C}$.  This is not obvious since the neutron-scattering measurement were performed  in zero magnetic field while the microwave measurements were performed  in finite field of  0.4 T to 1 T. Using Eq. (\ref{exchange})  we find the same value of the exchange integral, $J=$3.6 meV (for $S=$1.85 and $a=3.88$\AA) for the film and single crystal. This is also not obvious since thin films are strained. 
 
We processed in the same way the microwave absorption data of Ref. \cite{Dyakonov2,Dyakonov3} for the film of different composition -La$_{0.7}$Mn$_{1.3}$O$_{3}$ (see inset in  Fig. \ref{fig:M-D-all}).  While $D(M)$ proportionality holds at low temperatures, there is a an upward deviation as  $T\rightarrow T_{C}$. This suggests discontinuous variation of $D$ across the ferromagnetic transition. Our data for the La$_{0.7}$Sr$_{0.3}$MnO$_{3}$ films do not indicate such discontinuity.

\subsubsection{Linewidth} 
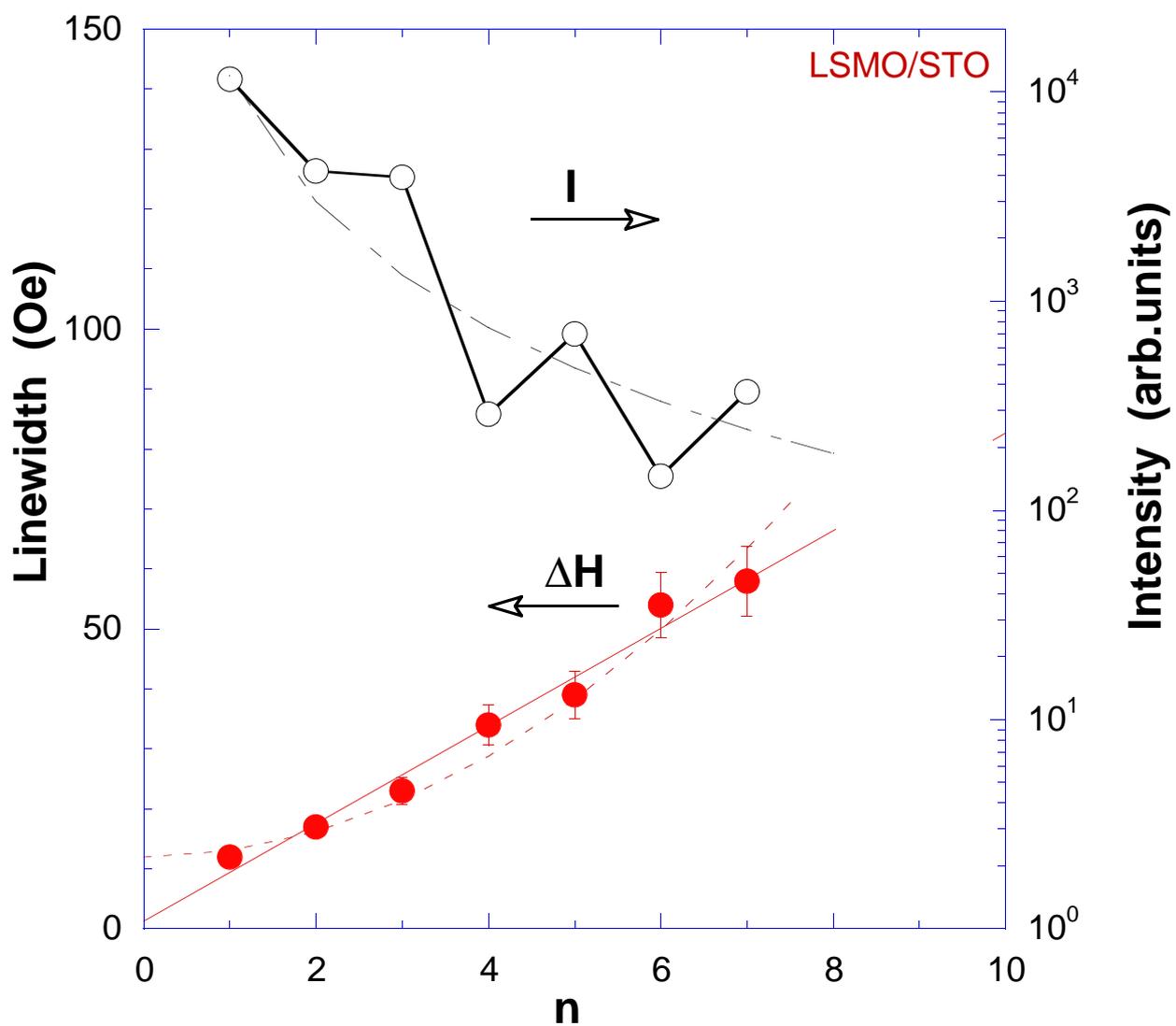
\begin{figure}[ht]
\caption{The filled symbols indicate the SWR linewidth,  $\Delta H_{n}$, the red solid line shows linear fit [Eq. (\ref{sc})], and the red dashed line shows quadratic fit [Eq. (\ref{linewidth-thickness})]. The open symbols indicate integrated SWR intensity corrected for the linewidth, $I=I_{pp}(\Delta H)^{2}$. Here, $I_{pp}$ is the peak-to-peak magnitude of the absorption derivative. The black solid line here is the guide to the eye, the dashed black line shows Kittel's $1/n^{2}$  prediction \cite{Kittel}.  The sample is a  200 nm thick La$_{0.7}$Sr$_{0.3}$MnO$_{3}$ film on the SrTiO$_{3}$ substrate at ambient temperature. }
\label{fig:linewidth-k}
\end{figure}

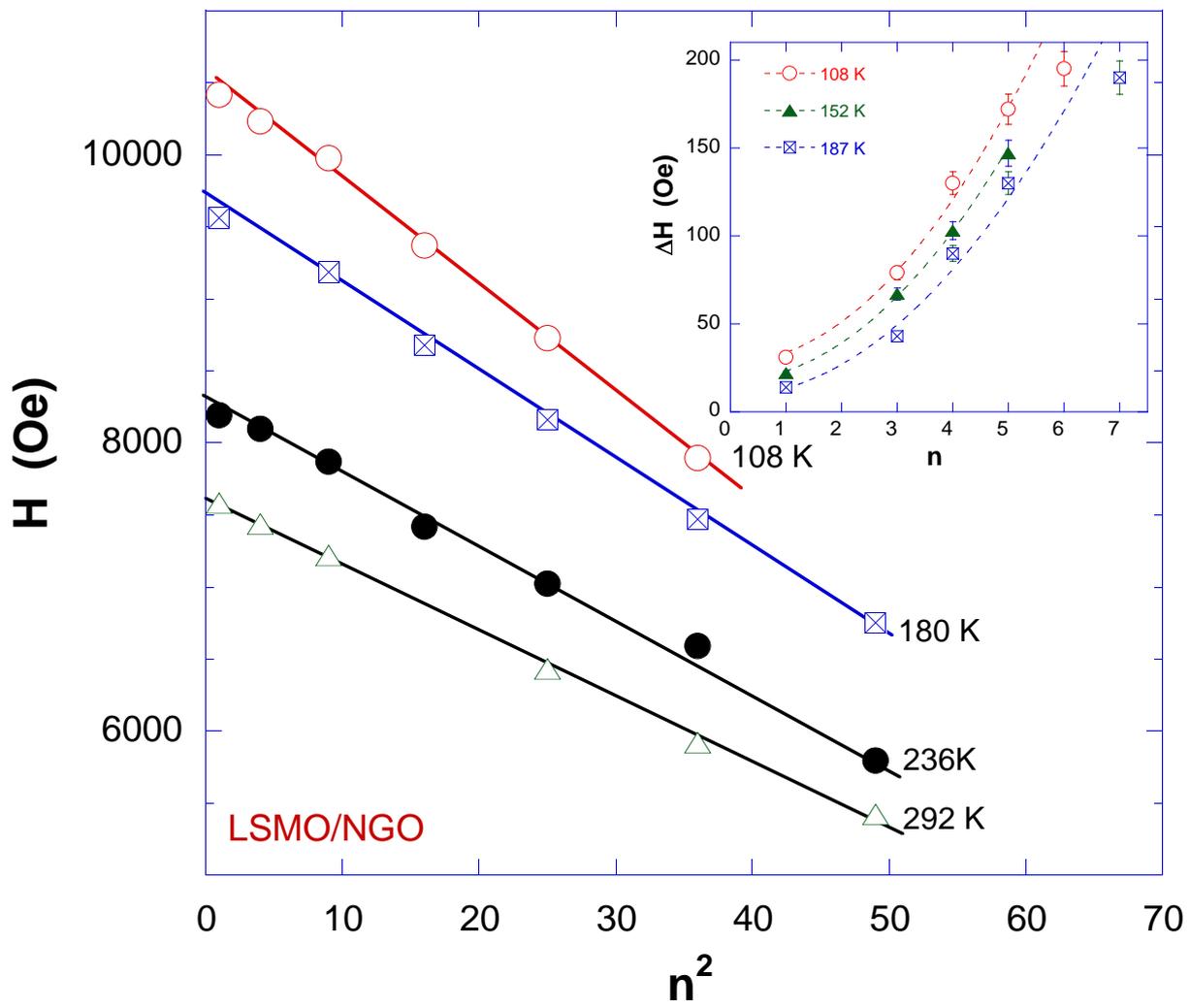
\begin{figure}[ht]
\caption{Resonance field of the SWR modes at several temperatures for the sample shown in Fig. \ref{fig:NGO-H}. The lines show $H_{n}\propto n^{2}$ approximation. The inset shows linewidth vs mode number. The dashed lines in the inset show quadratic approximation [Eq. (\ref{linewidth-thickness})].}
\label{fig:NGO-harm}
\end{figure}

The linewidth that steadily increases with mode number (Figs. \ref{fig:linewidth-k},\ref{fig:NGO-harm}) is not frequently observed in microwave absorption studies of spin-wave resonances. Previous studies of permalloy films and other highly conducting  ferromagnets  found weak and nonmonotonous dependence $\Delta H(n)$. (Refs. \cite {Lubitz,Suran,Frait,Bailey,Phillips2}) Such nonmonotonous dependence arises from the sum of several sources such as eddy-current damping, surface roughness, and fluctuations (exchange energy, anisotropy field, thickness). \cite{Levy}

   The $k$-dependent linewidth observed in our studies could be hardly intrinsic. Indeed, theoretical prediction for the magnon-magnon scattering in manganites yields inverse spin lifetime $\Gamma\propto k^{4}$ (Ref. \cite{Golosov}). The neutron scattering studies  of La$_{0.85}$Sr$_{0.15}$MnO$_{3}$ single crystals \cite{Vasiliu3} indeed yield $\Gamma=3050k^{4}$ (here $k$ is in \AA$^{-1}$ and $\Gamma$ is in meV). Extrapolation of this $k^4$ dependence (obtained for 10 K and $k=$0.02-0.16\AA$^{-1}$) down to the range of wavevectors  occurring in our microwave studies ($k_{max}=0.014\AA ^{-1}$) yields $\Gamma= 1.1\times 10^{-4}$. This is much smaller than what we observe in Fig. \ref{fig:linewidth-k} - $\Gamma= \Delta H/\gamma=4.7\times 10^{-3}$ (note however that our films have  different composition as compared to those  studied in Ref. \cite{Vasiliu3}). This indicates that the $k$- dependent linewidth found in our studies in thin films is extrinsic. It most probably arises either from the inhomogeneous broadening associated with  thickness nonuniformity, \cite{Bailey} or from the spin-wave scattering on surface and bulk disorder. \cite{Furukawa,Hoeppe,Scotter} 

If we take an extreme approach and assume only  thickness  
nonuniformity, then Ref. \cite{Bailey} yields
\begin{equation}
\Delta H_{d}=\Delta H_{0}+\frac{2D\pi^{2} n^{2}}{d^{3}}\Delta d,
\label{linewidth-thickness}
\end{equation}
where $\Delta H_{0}$ is the linewidth of the uniform precession mode and $\Delta d$ is the average thickness variation. Analysis of our  data  according to Eq. (\ref{linewidth-thickness}) (dashed lines in Figs. \ref{fig:linewidth-k},\ref{fig:NGO-harm}) yields $\Delta d$= 4 nm (2$\%$) and 6 nm (4$\%$) for the films on STO and on NGO, correspondingly.

If we go to another extreme and assume spin-wave scattering to be the only source of line broadening, then according to the transit-time treatment \cite {Scotter} we find 
\begin{equation}
\Delta H_{sc}-\Delta H_{0}=\frac{v_{g}}{\gamma l_{sc}}=\frac{2D\pi n}{\gamma dl_{sw}}
\label{sc}
\end{equation} 
where $v_{g}=2Dk$ is the group velocity and $l_{sw}$ is the mean free path.  Spin-wave scattering on bulk disorder is characterized by the combination of linear and quadratic $k$-dependence, \cite{Furukawa,Hoeppe} whereas the linear $k$-dependence implies constant mean free path. Analysis of our data according to Eq. (\ref{sc}) and assuming constant $l_{sw}$, yields $l_{sw}/d=$ 2 for the LSMO/STO film at ambient temperature (Fig. \ref{fig:linewidth-k}). For the LSMO/NGO film (Fig. \ref{fig:NGO-harm}) we find  $l_{sw}/d=$ 1.4 and 1.3 at 108 K and at 187 K, correspondingly.  The spin-wave mean-free-path being on the order of film thickness,  suggests that the scattering occurs predominantly at film interfaces. Such scattering may result from surface roughness \cite{Rezende} or from the localized Mn$^{4+}$-ions \cite{NMR} or Mn$^{2+}$-ions \cite{Mn-surface} at  film surface.

The observed $k$-dependence of the linewidth is most probably due to both mechanisms: thickness nonuniformity and scattering. Note however, that both extreme approaches [Eq. (\ref{linewidth-thickness}) and Eq. (\ref{sc})] yield long mean free path that practically excludes bulk scattering. This is surprising, since the stoichiometry of La$_{0.7}$Sr$_{0.3}$MnO$_{3}$ suggests two possible states of Mn-ion: Mn$^{3+}$ with spin $S$=3/2 and Mn$^{4+}$ with spin $S=$2. If  La$_{0.7}$Sr$_{0.3}$MnO$_{3}$ were Heisenberg ferromagnet this would certainly result in magnetic inhomogeneity. However, observation of standing spin-wave resonances with high mode numbers indicates negligible bulk scattering and a single magnetic state of Mn-ions in the bulk, as was already established by the NMR studies of Mn$^{55}$  in La$_{0.67}$Sr$_{0.63}$MnO$_{3}$ (Ref. \cite{NMR}).

\subsubsection{Other manganite compounds}   The standing spin-wave resonances have been observed so far in the manganite compounds of the general formula La$_{1-x}$A$_{x}$MnO$_{3}$ where A is Sr,Ca,Ba,Mn and $x\sim 0.3$ (Refs. \cite{Lofland,Dyakonov2,Dyakonov3,Aleshk2,Yin,Shames,Lyfar} and present work). The $x\sim 0.3$  composition corresponds to highest conductivity and is the most distant from phase boundaries, \cite{Imada} in other words, the phase separation effects should be least insignificant here. This composition has also the highest spin-wave stiffness \cite{Vasiliu2,Aleshk1} that indicates increased length of the exchange interaction. The very fact that so far there is no indication of the standing spin-wave resonances in  manganite compounds with $x\neq 0.3$ (we also didn't find any spin-wave resonances in  La$_{0.8}$Sr$_{0.2}$MnO$_{3}$ films) may serve as an indirect evidence of strong spin-wave scattering there.

\section{Conclusions}
We measured temperature dependence of the spin-wave stiffness in thin La$_{0.7}$Sr$_{0.3}$MnO$_{3}$ films as determined from  the standing spin-wave resonances at microwave frequencies. At ambient temperature, the spin-wave stiffness of thin films and of single crystals is the same, while at low temperatures the spin-wave stiffness of thin films is enhanced with respect to that of a single crystal.  

The spin-wave linewidth in our films is limited by the scattering at film interfaces. The very fact that we are able to observe spin-wave resonance in La$_{0.7}$Sr$_{0.3}$MnO$_{3}$ up to eighth order implies high degree of coherence  and very low bulk spin-wave scattering.   This is important for spintronics and means that La$_{0.7}$Sr$_{0.3}$MnO$_{3}$ can be used as a  spin conserving component.

\begin{acknowledgments}
We are grateful to Denis Golosov for illuminating discussions, to Xiangzhen  Xu  for the help with the handling of the samples, to Oscar Arnache who initiated these experiments, and to anonymous referee for instructive comments. 
\end{acknowledgments}
\pagebreak 

\begin{thebibliography}{}
\bibitem{temperature} Our measurements were performed with the thermocouple mounted at some distance from the sample. Since  thermalization is slow, to restrict the duration of the experiment to several hours, in some experimental runs we slowly heated the sample from low temperature and measured the spectra in a continuous way without waiting for thermalization. In such experiments  we determine only $D$ and $H_{a}$.
\end{thebibliography}
\end{document}